\documentclass[final]{siamart250211}

\usepackage{amsmath,amssymb,amsfonts}
\usepackage{braket}
\usepackage[per-mode=repeated-symbol]{siunitx}
\usepackage{dutchcal}
\usepackage{mathtools}
\usepackage[caption=false]{subfig}
\usepackage[nonumberlist]{glossaries}
\usepackage{algpseudocode}
\usepackage{booktabs}
\usepackage[final]{microtype}
\usepackage{longtable}
\usepackage[all]{nowidow}

\newsiamremark{remark}{Remark}
\newsiamremark{hypothesis}{Hypothesis}
\crefname{hypothesis}{Hypothesis}{Hypotheses}
\newsiamthm{claim}{Claim}

\newacronym{HPC}{HPC}{high performance computing}
\newacronym{LRZ}{LRZ}{Leibniz Supercomputing Center}
\glsdisablehyper

\newcommand{\maintitle}{Utilizing Partitioning for Distributed Tensor Network Contraction}

\DeclareMathOperator{\edge}{\texttt{edges}}
\DeclareMathOperator{\legs}{\texttt{legs}}
\DeclareMathOperator{\leaves}{\texttt{leaves}}
\DeclareMathOperator{\treeroot}{\texttt{root}}
\DeclareMathOperator{\parent}{\texttt{parent}}
\DeclareMathOperator{\children}{\texttt{children}}
\DeclareMathOperator{\shortestpath}{\texttt{path}}
\DeclareMathOperator{\objective}{\texttt{objective}}
\DeclareMathOperator*{\argmax}{arg\,max}

\DeclareMathOperator*{\parmin}{parallel\,min}

\DeclareMathOperator{\vc}{\texttt{vc}}

\DeclarePairedDelimiter\ceil{\lceil}{\rceil}

\newcommand{\N}{\mathbb{N}}
\newcommand{\C}{\mathbb{C}}
\newcommand{\ssep}{\mid}

\AddToHook{cmd/appendix/before}{
    \crefalias{section}{appendix}
}

\headers{Partitioning for Tensor Network Contraction}{M.~Geiger, Q.~Huang, C.~B.~Mendl}

\title{\maintitle\thanks{Submitted to the editors August 22, 2025.
\funding{We acknowledge the support and funding received for the MUNIQC-SC initiative under funding number 13N16191 from the VDI technology center as part of the German BMBF program.}}} 

\author{
Manuel Geiger\thanks{Department of Computer Science, CIT, Technical University of Munich, Garching, Germany 
  (\email{manuel.geiger@tum.de}, \email{keefe.huang@tum.de}, \email{christian.mendl@tum.de}).}
\and
Qunsheng Huang\footnotemark[2]
\and
Christian B.~Mendl\footnotemark[2]\phantom{\footnotesize 1}\thanks{Institute for Advanced Study, Technical University of Munich, Garching, Germany}
}

\ifpdf
\hypersetup{
  pdftitle={\maintitle},
  pdfauthor={M.~Geiger, Q.~Huang, and C.~B.~Mendl}
}
\fi

\begin{document}

\maketitle

\begin{abstract}
    Tensor networks have proven to be a valuable tool, for instance, in the classical simulation of (strongly correlated) quantum systems. As the size of the systems increases, contracting larger tensor networks becomes computationally demanding. In this work, we explore the use of partitioning to distribute the contraction task across a distributed memory architecture, commonly found in high-performance computing environments. This approach offers a promising alternative to the widely used parallelization based on slicing, which often incurs significant computational overhead. We particularly investigate how a given partitioning can be optimized to enable efficient contraction, as both computational and memory costs are highly sensitive to the chosen strategy. To this end, we introduce a simulated annealing-based method that iteratively refines a partitioning to minimize the operation count along the critical path, thereby reducing time-to-solution. The algorithm is evaluated on MQT~Bench circuits and achieves a lower computational cost in $47$ of $71$ circuits, with improvements up to $5$ orders of magnitude compared to contraction paths generated by Cotengra, a state-of-the-art contraction path finding library.
\end{abstract}



\section{Introduction}

Tensor networks provide a powerful framework for the classical simulation of quantum systems~\cite{Markov2008, Chen2018, Huang2020, Pednault2020,Berezutskii2025}.
This was demonstrated when they were used to disprove the first claim of quantum advantage by Google~\cite{Arute2019} by simulating the circuit in question on a small compute cluster within hours instead of the predicted $\num{10000}$ years~\cite{Pan2022a,Pan2022b}.
They were also used to refute a recent claim of quantum advantage by IBM~\cite{Kim2023}, showing that tensor networks can efficiently simulate a quantum algorithm deemed out of reach of classical simulators by the original authors~\cite{Patra2024}.
Tensor network methods have also been explored in other areas of research, such as machine learning, quantum optimization, quantum chemistry, or quantum error correction~\cite{Garcia2024,Berezutskii2025}.

Due to the high demands on computing power and memory, tensor network methods are fitting candidates for the field of \gls{HPC}.
There are multiple existing libraries for tensor networks utilizing \gls{HPC}, including QuantEx~\cite{QuantEx}, cuTensorNet~\cite{cuTensorNet}, ExaTN~\cite{ExaTN}, and Jet~\cite{Jet}.
Additionally, libraries for handling individual tensors or tensor operations include Taco~\cite{Taco}, \mbox{TBLIS}~\cite{TBLIS}, and cuTensor~\cite{cuTensor}.
Moreover, there exist multiple contraction-order finding algorithms, with the most ubiquitous implementations including opt\_einsum~\cite{Smith2018}, Cotengra~\cite{cotengra}, and flow-cutter~\cite{hamann_graph_2018}.
Most of these libraries use slicing to parallelize the contraction task.
Using partitioning of tensor networks for parallelization of the contraction is, to the best of our knowledge, unexplored.
However, tensor network partitioning has been used for contraction path finding~\cite{Gray2021, Pan2022a} and for finding a tree embedding~\cite{Ma2024}.

In this work, we investigate the use of tensor network partitioning to run the contraction task in a distributed-memory setting, introducing a cost function that considers contractions that can occur independently and simultaneously.
Furthermore, we employ simulated annealing to improve the initial partitioning of tensor networks, with the goal of optimizing the time-to-solution of the overall contraction operation.
We compare this with the HyperOptimizer available in the Cotengra library~\cite{Gray2021}, demonstrating a better average improvement of computational cost and memory cost.
Additionally, we propose a cost metric that incorporates parallelism by summing operations along the critical path, providing an estimate of time-to-solution.

\section{Definitions}\label{sec:definitions}
\subsection{Tensor Networks}
We define a tensor network as a mathematical graph-like data structure as follows. A tensor network is specified by a tuple $G = (V, E)$ consisting of a discrete set of tensor vertices $V$ and weighted tensor edges $E$.
Each vertex $v \in V$ is associated with a tensor $T_v \in \C^{n_{v,1} \times \cdots \times n_{v,d_v}}$, where $d_v \in \N_0$ denotes the \emph{degree} of the tensor and $n_v \in \N^{d_v}$ its \emph{dimensions}. Formally, we admit tensors of degree zero, which are scalars.
The set of tensor edges consists of tuple pairs:
\begin{align}
    E \subseteq \big\{ & ((u, a), (v, b)) \ssep{} u \in V, v \in V \cup {\{\bot\}},                               \\
                       & a \in \{ 1, \dots, d_u \}, b \in \{ 1, \dots, d_v \}, n_{u,a} = n_{v,b} \big\},\nonumber
\end{align}
where $\bot$ is a dummy vertex that indicates an unbound edge and $d_{\bot} = 1$.
We additionally define
\begin{equation}
    \edge(w) \coloneq \{((u,a),( v, b)) \in E \ssep{} u = w\,\vee\, v = w\}
\end{equation}
for some vertex $w \in V$.
The size $|T_v|$ of a tensor is the product of its dimensions,
\begin{equation}
    |T_v| = \prod_{i=1}^{d_v} n_{v,i} = \prod_{e \in \edge(v)} n(e),
\end{equation}
where $n(e),\,e \in E$ provides the dimension of an edge. For the sake of simplicity of notation in function definitions, we treat a tensor and its associated vertex as interchangeable, i.e., $\texttt{fun}(v) = \texttt{fun}(T_v)$.
\begin{figure}[htbp]
    \centering
    \includegraphics{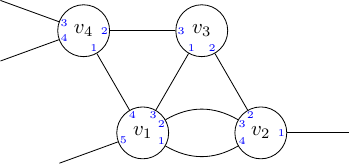}
    \caption{A general tensor network showing axes labels inside the individual tensor nodes.}
    \label{fig:general_tensor_network}
\end{figure}

\Cref{fig:general_tensor_network} illustrates a general tensor network diagram.
Note that several tensor axes are open (unbound), corresponding to the wires in the diagram attached to only one tensor.
Without loss of generality, we can assume that the tensor network graph is connected; a path, thus, exists between any pair of nodes. Otherwise, the subgraphs can be considered separately.

\subsection{Contractions}
Given such a setup, the task consists of contracting a selected set of bound edges in the network, resulting in an output network (possibly containing open axes).
For simplicity, we assume that the task is to contract \emph{all} bound edges in a network, resulting in a single tensor.
Additionally, we assume that all contractions occur in a pairwise fashion.

The contraction of two complex tensors $S \in \C^{\ell_1 \times \cdots \times \ell_f \times m_1 \times \cdots \times m_g}$ and $T \in \C^{m_1 \times \cdots \times m_g \times n_1 \times \cdots \times n_h}$ along the $g$ trailing axes of $S$ and $g$ leading axes of $T$ results in a tensor $R \in \C^{\ell_1 \times \cdots \times \ell_f \times n_1 \times \cdots \times n_h}$ with entries
\begin{equation}
    \label{eq:basic_contraction}
    R_{i_1, \dots, i_f, k_1, \dots, k_h} = \sum_{j_1, \dots, j_g} S_{i_1, \dots, i_f, j_1, \dots, j_g} T_{j_1, \dots, j_g, k_1, \dots, k_h}.
\end{equation}
Note the similarity to a matrix-matrix multiplication.
This definition can be generalized to permutations (transpositions) of the axes of $R$, $S$, and $T$, as long as the to-be contracted axes have pairwise identical dimensions.

\subsection{Contraction Trees}

Given the contraction task defined in the previous section, choosing a good ordering of contractions is crucial in improving time-to-solution~\cite{Gray2021,Ors2019}.
A critical insight is that the final result is independent of the order in which the pairwise contractions are performed.
Yet, the order determines the shape of \emph{intermediate tensors} and significantly influences the computational cost of intermediate contractions.
Optimal contraction orderings exist for tensor networks with a fixed structure, such as the well-known MPS formulation and its various associated algorithms, such as the TEBD or DMRG algorithms~\cite{White1992,verstraete_matrix_2008,schuch_computational_2008,PerezGarcia2007,van_damme_efficient_2021}.
However, finding a good ordering is NP-hard for general tensor networks, with possible solutions inhabiting an exponentially growing space~\cite{ChiChung1997,Markov2008}.
While the optimal contraction path can only be found via exhaustive search, heuristic algorithms exist to find good paths in practice.
Thus, extensive investigation has been done on \emph{heuristically} finding a good contraction ordering~\cite{Gray2021,Smith2018}.
For instance, the \emph{Greedy} heuristic constructs a contraction path by iteratively selecting the contraction that maximizes a specified objective function at each step, such as the reduction in memory caused by the contraction.

\begin{figure}[htbp]
    \centering
    \includegraphics{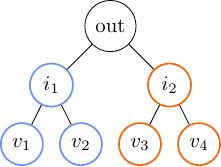}
    \caption{One possible contraction tree for the problem given in \cref{fig:general_tensor_network}. The left and right contractions can occur simultaneously.}
    \label{fig:general_contraction_tree}
\end{figure}

The order of contractions in a tensor network is canonically represented by a rooted binary tree, known as a \emph{contraction tree}.
For an arbitrary tensor network $G = (V, E)$, we define its tree embedding $(B, b)$ as follows:
\begin{itemize}
    \item $B$ is a binary tree $B = (\mathcal{V}_B, \mathcal{E}_B)$, where $\mathcal{V}_B$ represents a set of tree vertices and $\mathcal{E}_B$ consists of tree edges.
    \item $b: V \to \leaves(B)$ is a bijective mapping, associating each tensor $v \in V$ with a leaf node in $B$.
\end{itemize}
Hence, the leaf nodes of $B$, denoted as $\leaves(B)$, correspond directly to the tensors in $V$. The root of $B$, denoted as $\treeroot(B)$, is associated with the final output tensor of the network.
Furthermore, the parent of any two child nodes represents the intermediate tensor obtained by contracting the two tensors associated with the child nodes.
Contraction trees do not have redundant nodes, meaning that each parent node has exactly two children $(\mathcal{c}_1, \mathcal{c}_2) = \children(\mathcal{p})$, and each non-root vertex has exactly one parent $\mathcal{p} = \parent(\mathcal{c}_1) = \parent(\mathcal{c}_2)$, where $\mathcal{p}, \mathcal{c}_1, \mathcal{c}_2 \in \mathcal{V}_B$.
The set of tree edges is then defined as
\begin{equation}
    \mathcal{E}_B = \big\{ (\mathcal{v}, \parent(\mathcal{v})) \ssep{} \mathcal{v} \in \mathcal{V}_B \setminus \{\treeroot(B)\} \big\}.
\end{equation}
Additionally, we define the function $\shortestpath(\mathcal{v}, \mathcal{u})$ to reference the vertices on the path between tree vertices $\mathcal{v}$ and $\mathcal{u}$, as well as the function  $\legs(\mathcal{v})$ that returns the edges of the (intermediate) tensor associated with a tree vertex $\mathcal{v}$, being defined as:
\begin{align}
    \mathbf{I}(\mathcal{v}) & = \bigcup\limits_{c \in \children(\mathcal{v})} \legs(c) \setminus \bigcap\limits_{c \in \children(\mathcal{v})} \legs(c)\nonumber \\
    \legs(\mathcal{v})      & =
    \begin{cases}
        \edge(b^{-1}(\mathcal{v})), & \text{if } \mathcal{v} \in \leaves(B)      \\
        \mathbf{I}(\mathcal{v}),    & \text{if } \mathcal{v} \not\in \leaves(B),
    \end{cases}
\end{align}

We refer to any such binary tree $B$ as a contraction tree of $G$. It is important to note that multiple valid tree embeddings may exist for a given tensor network $G$, which correspond to an exponentially growing space of viable contraction paths to explore during optimization. An example of a contraction tree is depicted in \cref{fig:general_contraction_tree}.

\subsection{Partitioning}
\begin{figure}[htbp]
    \centering
    \includegraphics[scale=0.8]{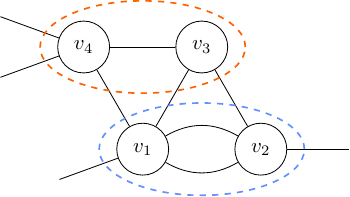}
    \hspace{2cm}
    \includegraphics[scale=0.8]{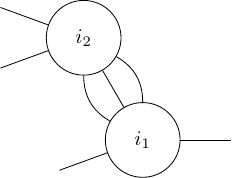}
    \caption{Left: The tensor network of \cref{fig:general_tensor_network} split into two partitions. Right: The two tensors resulting from contracting the partitions individually.}
    \label{fig:general_partitioning}
\end{figure}

For an arbitrary tensor network $G = (V, E)$, a partitioning $K$ of the vertices $V$ is considered valid if it satisfies the following conditions:
\begin{enumerate}
    \item $\bigcup_{k \in K} k = V$, meaning that the union of all partitions covers the entire vertex set,
    \item $\forall k_i, k_j \in K, k_i \neq k_j \implies k_i \cap k_j = \emptyset$, ensuring that the partitions are disjoint, and
    \item $\forall k \in K,\, k \neq \emptyset$, forbidding empty partitions.
\end{enumerate}
Given such a partitioning, a tree embedding $(B, b)$ of $G$ is said to accept this partitioning if it meets the following criterion:
\begin{equation}
    \forall k \in K,\ \exists \mathcal{v} \in \mathcal{V}_B:\
    \left\{ b^{-1}(\mathcal{u}) \ssep{} \mathcal{u} \in \leaves(B_{\mathcal{v}}) \right\} = k,
\end{equation}
where $B_{\mathcal{v}}$ denotes the subtree of $B$ rooted at $\mathcal{v}$.
In other words, for each partition, there is a subtree where the leaf nodes correspond to the tensors that form the partition.
For convenience, we use $B_k$ to refer to a subtree that satisfies this criterion for a given partition $k$.
\Cref{fig:general_partitioning} shows an example partitioning with $k_1 = \{v_1, v_2\}$, $k_2 = \{v_3, v_4\}$, and $K = \{ k_1, k_2 \}$. The tree embedding shown in \cref{fig:general_contraction_tree} accepts this partitioning since the leaves of $B_{i_1}$ and $B_{i_2}$ form the partitions $k_1$ and $k_2$, respectively.

\section{Problem Statement and Metrics}
\label{sec:metrics}
Now that we have defined the overall contraction problem, the goal is then to identify a partitioning whose corresponding contraction tree minimizes the time-to-solution compared to other state-of-the-art methods in a parallelized setting. In order to compare methods with different forms of parallelization, We approximate the time-to-solution with operation count along the critical path, by introducing the metric \cref{eq:con_par} as opposed to the standard cost, defined as \cref{eq:con_serial}. As most contraction operations are bounded my memory constraints, we additionally track this metric, defined in \cref{eq:mem}

\subsection{Sequential Contraction Cost}
In a serial setting, we can define the computational or contraction complexity as
\begin{equation}
    \texttt{con}_{\textnormal{serial}}(B) = \sum\limits_{\mathcal{v} \in \mathcal{V}_B \setminus \leaves(\mathcal{V}_B)} 2^{\vc(\mathcal{v})},
    \label{eq:con_serial}
\end{equation}
where $\vc$ is the \emph{vertex congestion}~\cite{OGorman2019}:
\begin{align}
    \mathbf{E}(\mathcal{v}) & =  \bigcup_{\mathcal{c} \in \children(\mathcal{v})} \legs(\mathcal{c}), \nonumber \\
    \vc(\mathcal{v})        & = \sum_{e \in \mathbf{E}(\mathcal{v})}\log_2\left( n(e) \right).
\end{align}

\subsection{Parallel Contraction Cost}

However, in the shared-memory scenario, where independent contractions are executed in parallel, the complexity of the parallel time-to-solution can be expressed as:
\begin{align}
    \mathbf{P}                         & = \big\{ \shortestpath \left(\mathcal{v}, \treeroot(\mathcal{V}_B) \right) \setminus \{ \mathcal{v} \} \ssep{} \mathcal{v} \in \leaves(\mathcal{V}_B) \big\}, \nonumber \\
    \texttt{con}_{\textnormal{par}}(B) & = \max_{p \in \mathbf{P}} \sum_{\mathcal{v} \in p} 2^{\vc(\mathcal{v})}.
    \label{eq:con_par}
\end{align}
This formulation calculates the contraction complexity along the \textit{critical path}, as described by \cite{OGorman2019}.
However, this cost function does not apply to the distributed case, where both local and inter-node parallelism need to be considered along with communication costs.

In a distributed-memory setting with a valid partitioning $K$, the contraction of partitions happens in parallel, followed by inter-node contractions with communication overhead; the breakdown of the steps is detailed in \cref{sec:methods}. The cost for this contraction task is defined as:
\begin{align}
    \shortestpath_k                        & = \shortestpath(\treeroot(B_k), \treeroot(B)) \setminus \{\treeroot(B_k)\}, \nonumber                                                                                                    \\
    \texttt{con}_{\textnormal{dist}}(B, K) & = \max_{k \in K}\left\{ \texttt{con}(B_k) + \sum_{\mathcal{v} \in \shortestpath_k } \left( 2^{\vc(\mathcal{v})} + \min_{c \in \children(\mathcal{v})} \texttt{comm}(c) \right) \right\},
    \label{eq:con_dist}
\end{align}
where \(\texttt{con}\) can be replaced with either \(\texttt{con}_{\textnormal{serial}}\) or \(\texttt{con}_{\textnormal{par}}\) depending on the usage of shared-memory parallelism.
The function \(\texttt{comm}(c)\) denotes the cost of communicating the tensor associated with the node $\mathcal{c}$.
We assume that only one tensor needs to be communicated, and naturally, the smaller is chosen.
If communication costs are not considered and intra-node parallelism is used, we then recover \cref{eq:con_par}.
For our experiments, communication costs are not included in theoretical calculations as empirical results indicate that these costs are far smaller than the tensor contraction costs.

Optimizing this structure to reduce the overall contraction cost has been studied before, with e.g. \cite{Ibrahim2022} utilizing a dynamic programming approach. However, to the best of our knowledge, utilizing a cost function that reduces time-to-solution in a distributed use case has not been investigated. In this work, the introduced algorithms attempt to identify contraction trees $B$ that minimize \cref{eq:con_dist}.

\subsection{Memory Cost}
While we utilize \cref{eq:con_dist} as the main cost to minimize in this work, the memory requirements are often a limiting factor when running on hardware. Thus, we additionally track and compare the spatial requirements of the resultant contraction trees. We define the memory cost of a contraction task as the maximum memory required for any single contraction operation in the tree (up to a constant depending on data type):
\begin{multline}\label{eq:mem}
    \texttt{mem}(B) = \max_{\mathcal{v} \in \mathcal{V}_B \setminus \leaves(B)} \Bigg\{ \prod_{\ell \in \legs(\mathcal{v})} n(\ell) + \prod_{\ell_1 \in \legs(\mathcal{c}_1)} n(\ell_1) + \prod_{\ell_2 \in \legs(\mathcal{c}_2)} n(\ell_2)\ssep{}\\(\mathcal{c}_1, \mathcal{c}_2)= \children(\mathcal{v})\Bigg\}.
\end{multline}

This naively assumes that intermediate tensors not relevant to the current computation can be stored on disk in the worst case.
However, alternative memory cost calculations for the used hardware can be introduced without loss of generality.

Note that \cref{eq:mem} assumes that only one contraction happens in parallel. For parallel contractions in a shared-memory system, the maximum required memory depends on the timing of the instructions, which can vary at run time.

\section{Methods}\label{sec:methods}
Naively, given the task of contracting an arbitrary tensor network in a distributed manner, we can split the task into five steps:
\begin{enumerate}
    \item Partitioning the tensor network.
    \item Determining a communication or reduction path that specifies how tensors are communicated for inter-partition or inter-node contractions.
    \item Distributing the partitions across nodes in the \gls{HPC} system.
    \item Fully contracting the partitions locally.
    \item Performing the inter-node contraction or fan-in operation following the path from step 2, contracting the resultant local tensors on each node.
\end{enumerate}
Steps 2 and 4 are the realization of the contraction tasks described in \cref{sec:definitions} and have been widely investigated in the literature.
We give a short note about how we implemented these steps in \cref{ssec:partitioning_score}.
The main focus, however, lies on improving step 1, which is unique to the distributed \gls{HPC} setting and crucial for contraction performance.
In particular, we investigate how simulated annealing can be used to refine an initial partitioning for optimizing time-to-solution.

\subsection{Slicing}
Before looking at partitioning, we note that many existing methods employ \emph{slicing} or Feynman contraction methods to parallelize the contraction task~\cite{Pan2022a, Chen2018, Gray2021, ExaTN, Huang2020, Jet}.
Slicing a tensor leg of dimension $n(\ell)$ entails performing $n(\ell)$ separate contractions, each corresponding to a fixed value of the sliced index $\ell$.
The resulting $n(\ell)$ tensors are subsequently summed together to yield the final output tensor.
When $k$ legs are sliced, the total number of tensor network contractions required grows multiplicatively as $\prod_i^k n(\ell_i)$.
The individual slices can be contracted embarrassingly parallel; hence, slicing is a prime candidate for parallelization.
However, slicing introduces computational overhead due to redundant operations across the multiple contractions.
In contrast, partitioning does not incur a computational overhead, which motivates our focus on partitioning-based techniques in this work.

\subsection{Initial Partitioning}
We assume that the number of compute nodes equals the number of partitions.
Having more nodes than partitions would leave some nodes underutilized, whereas having more partitions than nodes would require executing multiple partitions sequentially on the same node.
We also assume that the number of required partitions is known a priori.
Then, an initial partitioning of the tensor network can be found. In \cite{Gray2021}, the KaHyPar partitioning library is utilized with settings that optimize for:
\begin{itemize}
    \item Reducing the size of intermediate tensors after local contraction.
    \item Keeping the average partition size similar based on a chosen imbalance parameter.
\end{itemize}
Reducing the size of intermediate tensors directly reduces communication costs, which benefits the time-to-solution in the distributed case.
However, these chosen settings do not directly correlate to load balancing from a parallelized contraction cost perspective.
Hence, in this work, we propose to include an additional balancing step to improve the partitions from a load-balancing perspective, using the existing partitioning scheme as a starting point.

\subsection{Simulated Annealing Partitioning Refinement}
\begin{figure}[htbp]
    \centering
    \includegraphics{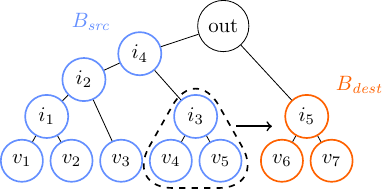}
    \hfill
    \includegraphics{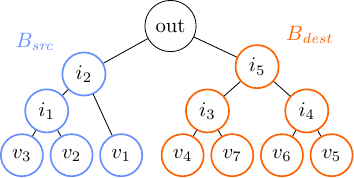}
    \caption{Example of shifting the intermediate tensor $i_3$, or equivalently, shifting the leaf tensors, $v_4$, $v_5$, belonging to a subtree in partition $B_{\textnormal{src}}$ to a different partition $B_{\textnormal{dest}}$. The contraction tree structure is not preserved in the move.}
    \label{fig:rebalancing}
\end{figure}

To avoid the exponential cost of an exhaustive search, we attempt to find a better partitioning via local updates: namely shifting tensors from the contraction tree of the source partition $B_{\mathit{src}}$ to that of the destination partition $B_{\mathit{dest}}$.
An example is shown in \cref{fig:rebalancing},  where a small subtree rooted at $i_3$ is shifted from partition $B_{src}$ to partition $B_{dest}$, redistributing the load between two unbalanced partitions.
Understanding that the term subtree can refer to both intermediate tensors and full partitions, we shall, henceforth, only use it to refer to the former.

The problem then is to identify a good subtree or intermediate tensor to shift and a good partition to move it to.
Identifying these local updates is at the heart of our simulated annealing algorithm used for partition improvement.
We note that simulated annealing has already been used previously to optimize contraction paths~\cite{Kalachev2021}, but without focus on parallelization.

\subsubsection{Simulated Annealing}
The two major difficulties in optimizing contraction trees are the super-polynomial search space and the sensitivity of the problem to perturbation.
There is an exponential number of possible shifts between partitions, and moving even a single tensor can result in wildly different contraction costs.
Hence, we apply a probabilistic approach to explore this exponential space.

Simulated annealing~\cite{Kirkpatrick1983,Cerny1985} is an optimization technique for finding the global minimum of a function.
At each step, the algorithm selects a subtree to shift and identifies a partition to shift it to.
This effectively identifies a ``neighbor'' of the current contraction tree or \emph{state} in the search space, which is separated from the current state by a single shift.

The cost functions identified in \cref{sec:metrics} are used to calculate the associated cost of each state.
If the neighbor's cost is lower, the algorithm moves to this state as a definite improvement is identified.
If the neighbor's cost is higher, the algorithm moves with probability based on the difference between the two costs and a temperature factor, which slowly decreases with each iteration; the longer the algorithm runs, the less likely a state with a worse cost is accepted.
This probabilistic acceptance of higher-cost neighbors allows the algorithm to escape local minima.
The usual acceptance probability function for the second case is
\begin{equation}
    P(c, c_\mathit{new}, T) = \exp\left(-\frac{c_\mathit{new} - c}{T}\right),
\end{equation}
where $c$ and $c_{\text{new}}$ are the current and new cost and $T$ is the temperature~\cite{Kirkpatrick1983}.
Each temperature iteration can comprise several steps, meaning the same temperature can be used for multiple subsequent update steps.

\subsubsection{Algorithm Modifications}
The initial and final temperatures are problem-specific parameters.
Since we use the total operational cost as the score of a partitioning, as detailed in \cref{ssec:partitioning_score}, the temperatures would need to be tuned individually for each tensor network if the usual acceptance probability function is used.
Instead, we use the logarithm of the contraction costs, turning the acceptance probability function into
\begin{equation}
    P(c, c_\mathit{new}, T)= \exp\left(-\frac{\log\left(\frac{c_\mathit{new}}{c}\right)}{T}\right),
\end{equation}
hence making the cost difference relative.

Additionally, we modify the algorithm to work with a given time limit rather than a fixed number of temperature iterations.
For this, we compute the temperature at each iteration by interpolating the initial and final temperatures, based on the proportion of elapsed time.
The cooling schedule is smooth since the duration of individual temperature steps is approximately uniform in practice.
We employ an exponential cooling schedule (originally introduced in \cite{Kirkpatrick1983}), with the temperature $T$ at some time fraction $t \in [0, 1]$ being computed as
\begin{equation}
    T(t) = T_0 \cdot \alpha^{t},
\end{equation}
where $T_0$ is the initial temperature and $\alpha$ is a constant factor that we choose as
\begin{equation}
    \alpha = \frac{T_f}{T_0},
\end{equation}
such that we arrive at a given final temperature $T_f$ for $t = 1$.
This schedule yielded slightly better results in practice than a linear cooling schedule.

Furthermore, we parallelize the simulated annealing using the division algorithm~\cite{Aarts1990}.
Instead of executing $N$ steps per temperature iteration on a single processor, we perform $\ceil{N/p}$ steps independently on $p$ processors.
Then, the best solution among the processors is selected as the starting point for the next temperature iteration.

Moreover, we keep track of the best solution found at any temperature iteration.
If no better solution is found for a given number of iterations, we restart from the hitherto best solution.
In the end, the best solution found is returned. The general algorithm with all mentioned modifications is shown in \cref{alg:sa}.
\begin{algorithm}
    \begin{algorithmic}
        \Function{DoSteps}{$N, s_\mathit{current}, c_\mathit{current}, T$}
        \For{$j = 1$ \textbf{to} $N$}
        \State $s_\mathit{new} \coloneq \Call{SelectNeighbor}{s_\mathit{current}}$
        \State $c_\mathit{new} \coloneq \Call{EvaluateState}{s_\mathit{new}}$
        \State $P_\mathit{accept} = \exp(-\log(c_\mathit{new}/c_\mathit{current}) / T)$
        \If{$P_\mathit{accept} \ge \Call{Random}{0, 1}$}
        \State $s_\mathit{current}, c_\mathit{current} \coloneq s_\mathit{new}, c_\mathit{new}$
        \EndIf
        \EndFor
        \State \textbf{return} $s_\mathit{current}, c_\mathit{current}$
        \EndFunction
        \Statex
        \Function{SimulatedAnnealing}{$T_0, T_f, N_\mathit{steps}, s_\mathit{initial}, \mathit{time\_limit}$}
        \State $s_\mathit{best} \coloneq s_\mathit{current} \coloneq s_\mathit{initial}$
        \State $c_\mathit{best} \coloneq c_\mathit{current} \coloneq \Call{EvaluateState}{s_\mathit{initial}}$
        \State $i_\mathit{best} \coloneq i \coloneq-1$
        \State Compute cooling factor $\alpha \coloneq T_f / T_0$
        \State Compute steps per processor $N_p \coloneq \ceil{N_\mathit{steps} / \#\mathit{Processors}}$
        \State $\mathit{elapsed} \coloneq 0$
        \While{$\mathit{elapsed} < \mathit{time\_limit}$}
        \State Increment iteration $i \coloneq i + 1$
        \State Compute time progress $t \coloneq \mathit{elapsed} / \mathit{time\_limit}$
        \State Set temperature $T \coloneq T_0 \cdot \alpha^t$
        \State $s_\mathit{current}, c_\mathit{current} \coloneq \parmin \Call{DoSteps}{N_p, s_\mathit{current}, c_\mathit{current}, T}$
        \If{$c_\mathit{current} < c_\mathit{best}$}
        \State $i_\mathit{best}, s_\mathit{best}, c_\mathit{best} \coloneq i, s_\mathit{current}, c_\mathit{current}$\Comment{Update best solution}
        \ElsIf{$i - i_\mathit{best} \ge$ RestartThreshold}
        \State $i_\mathit{best}, s_\mathit{current}, c_\mathit{current} \coloneq i, s_\mathit{best}, c_\mathit{best}$\Comment{Restart from best solution}
        \EndIf
        \State Update $\mathit{elapsed}$ time
        \EndWhile
        \State \textbf{return} $s_\mathit{best}$
        \EndFunction
    \end{algorithmic}
    \caption{Simulated annealing algorithm.}
    \label{alg:sa}
\end{algorithm}

\subsubsection{Subtree and Partition Selection}
\begin{algorithm}
    \begin{algorithmic}
        \Function{SelectTargetNaive}{$K, k_\mathit{src}$}
        \State \textbf{return} random $k_\mathit{dest}$ from $K \setminus \{k_\mathit{src}\}$
        \EndFunction
        \Statex
        \Function{SelectTargetDirected}{$K, k_\mathit{src}, \mathcal{v}, (B, b)$}
        \State \textbf{return} $\argmax_{k \in K \setminus \{k_\mathit{src}\}} \objective(\legs(\mathcal{v}), \legs(\treeroot(B_k)))$
        \EndFunction
        \Statex
        \Procedure{Rebalance}{$K, (B, b)$}
        \State Select random $k_\mathit{src}$ from $K$
        \State Select random tree node $\mathcal{v}$ from $B_{k_\mathit{src}}$
        \State Get associated tensors $T \coloneq b^{-1}(\leaves(B_{k_\mathit{src}}))$
        \State Get $k_\mathit{dest}$ using $\Call{SelectTargetNaive}{}$ or $\Call{SelectTargetDirected}{}$
        \State Move all $t \in T$ from $k_\mathit{src}$ to $k_\mathit{dest}$
        \State Update $B$\Comment{Find new contraction paths for the changed partitions, find new reduction path.}
        \EndProcedure
    \end{algorithmic}
    \caption{Rebalancing step of the two rebalancing models.}
    \label{alg:rebalancing}
\end{algorithm}

The final component is the methodology to select a neighbor of the current state, that is, selecting and applying a local update to the current partitioning.
We introduce two selection functions, \textsc{SelectTargetNaive} and \textsc{SelectTargetDirected}, detailed in pseudo-code in \cref{alg:rebalancing}.
Both functions start by randomly selecting a subtree of a randomly selected source partition.
This identifies the intermediate tensor that is shifted between partitions; we only allow the movement of subtrees that do not empty a given partition.
Then, a target partition is selected by the following means:
\begin{itemize}
    \item
          \textsc{SelectTargetNaive} selects a random target partition.
    \item
          \textsc{SelectTargetDirected} uses an objective function to find the best matching target partition for the tensors to be moved.
          We use the common Greedy heuristic for minimizing resultant memory as an objective function to estimate the improvement when shifting tensors:
          \begin{equation}
              \objective(T_\mathit{src}, T_\mathit{dest}) = |T_\mathit{src}| + |T_\mathit{dst}| - |T_{\mathit{result}}|,
          \end{equation}
          where $T_\mathit{src}$ and $T_\mathit{dest}$ are tensors in $B_\mathit{src}$ and $B_\mathit{dest}$, respectively, and $T_\mathit{result}$ is the tensor resultant from contracting these two tensors.
\end{itemize}
Then, we shift the leaf tensors of the selected subtree to the target partition and identify new contraction and reduction paths for the updated partitions.
The pseudo-code for the update step is shown in \cref{alg:rebalancing}.

\subsection{Cost Calculation}\label{ssec:partitioning_score}
Given an existing partitioning, we use a Greedy approach to find a contraction tree for each partition.
With this, we apply the cost functions \cref{eq:con_serial} or \cref{eq:con_par} to obtain the contraction cost per partition, which estimates the time needed to perform the intra-node contraction.
Then, we identify the inter-node communication.
Since the partition tensors can be large, investing more time in finding a good reduction path is advisable.
For this, we use a variant of the Greedy heuristic called RandomGreedy, which samples many random Greedy paths and selects the best one.
Given the reduction path, an estimate of the overall time-to-solution can be subsequently obtained using \cref{eq:con_dist}.
This estimate is the cost of the state that we try to minimize.
While we utilize a naive contraction order finder, more complex variants can be used similarly in future exploratory work.

\newpage
\section{Implementation and Experiments}
\subsection{Implementation Details}
Given the algorithm in \cref{alg:rebalancing}, we utilize KaHyPar as a starting point for partitioning the tensor network using a min-cut heuristic and default imbalance parameters\footnote{\raggedright We utilize the base settings available in the KaHyPar repository, specifically \textit{cut\_kKaHyPar\_sea20.ini}~\cite{Andre2018}}.
The contraction paths are found using Cotengra using the Greedy and RandomGreedy heuristics.
We realize the contractions following \cref{eq:basic_contraction} by two permutations and a matrix-matrix multiplication.
For this, we utilize the HPTT library~\cite{Springer2017} for the permutation of the data and MKL~\cite{MKL} for the multiplication.
We perform all contractions of a partition in serial due to memory limits, using all cores for the single matrix-matrix multiplications.
For communication, we use blocking point-to-point communication via MPI~\cite{MPI}.
Our code is publicly available\footnote{\raggedright \url{https://github.com/qc-tum/TNC}}.

\subsection{Test Setup}
In order to examine different circuits across multiple use cases, we utilized a series of circuits from the MQT~Bench benchmarks~\cite{Quetschlich2023}, which offers a unified suite of benchmark algorithms.
For circuits with configurable sizes, we chose instances with $10$, $30$, and $50$ qubits.
The circuits we used are listed in \cref{tab:mqtcircuits}.
For each circuit, we evaluated a single output amplitude, namely the amplitude of the $\ket{0\ldots0}$ basis state.
Each $N$-qubit gate was translated to the corresponding tensor of degree $2^{2N}$, and tensors for the bras and kets were added.
The quantum circuits and corresponding tensor networks were not simplified prior to contraction.

\begin{table}[htbp]
    \caption{List of MQT~Bench circuits used as benchmark.}
    \label{tab:mqtcircuits}
    \small
    \centering
    \begin{tabular}{lr@{\hskip 0.5ex}r@{\hskip 0.5ex}rr@{\hskip 0.5ex}r@{\hskip 0.5ex}r}\toprule
        \textbf{Name}       & \multicolumn{3}{c}{\textbf{\#Qubits}} & \multicolumn{3}{c}{\textbf{\#Tensors}}                            \\ \midrule
        ae                  & 10,                                   & 30,                                    & 50 & 255, & 2265, & 6275 \\
        dj                  & 10,                                   & 30,                                    & 50 & 48,  & 148,  & 248  \\
        ghz                 & 10,                                   & 30,                                    & 50 & 30,  & 90,   & 150  \\
        graphstate          & 10,                                   & 30,                                    & 50 & 40,  & 120,  & 200  \\
        groundstate\_large  &                                       &                                        & 14 &      &       & 252  \\
        groundstate\_medium &                                       &                                        & 12 &      &       & 192  \\
        groundstate\_small  &                                       &                                        & 4  &      &       & 32   \\
        grover-v-chain      &                                       &                                        & 11 &      &       & 6926 \\
        portfolioqaoa       & 10,                                   & 13,                                    & 17 & 465, & 780,  & 1326 \\
        portfoliovqe        & 10,                                   & 14,                                    & 18 & 195, & 357,  & 567  \\
        pricingcall         & 5,                                    & 15,                                    & 25 & 142, & 938,  & 8510 \\
        pricingput          & 5,                                    & 15,                                    & 25 & 142, & 956,  & 8536 \\
        qaoa                & 10,                                   & 13,                                    & 16 & 110, & 143,  & 176  \\
        qft                 & 10,                                   & 30,                                    & 50 & 270, & 2310, & 6350 \\
        qftentangled        & 10,                                   & 30,                                    & 50 & 280, & 2340, & 6400 \\
        qnn                 & 10,                                   & 30,                                    & 50 & 339, & 2819, & 7699 \\
        qpeexact            & 10,                                   & 30,                                    & 50 & 266, & 2331, & 6396 \\
        qpeinexact          & 10,                                   & 30,                                    & 50 & 276, & 2336, & 6396 \\
        qwalk-v-chain       &                                       & 11,                                    & 21 &      & 1595, & 5875 \\
        random              &                                       & 10,                                    & 30 &      & 403,  & 4685 \\
        realamprandom       & 10,                                   & 30,                                    & 50 & 195, & 1485, & 3975 \\
        routing             & 2,                                    & 6,                                     & 12 & 15,  & 51,   & 105  \\
        su2random           & 10,                                   & 30,                                    & 50 & 195, & 1485, & 3975 \\
        tsp                 & 4,                                    & 9,                                     & 16 & 47,  & 112,  & 203  \\
        twolocalrandom      & 10,                                   & 30,                                    & 50 & 195, & 1485, & 3975 \\
        vqe                 & 10,                                   & 13,                                    & 16 & 68,  & 89,   & 110  \\
        wstate              & 10,                                   & 30,                                    & 50 & 57,  & 177,  & 297  \\
        \bottomrule\end{tabular}
\end{table}

In each experiment, for each circuit, we ran our algorithms with a different number of partitions and selected the partition number that performed the best.
More specifically, the number of partitions per problem was chosen a priori by a sweep over the values $\{4, 8, 16, 32, 64, 128, 256\}$.
All methods were given $10$ minutes for finding a contraction path.
To account for the inherent randomness in the methods, we averaged the results of two runs per circuit.
Since we only considered inter-node parallelism for the experiments, we used \cref{eq:con_dist} for the cost calculation to determine the inter-node reduction and \cref{eq:con_serial} for the intra-node calculations for methods involving partitioning.

For comparison with the state of the art, we benchmarked the performance of the introduced algorithms against the HyperOptimizer method found in the Cotengra library.
This method finds a contraction path while slicing legs, leading to an alternative parallelization strategy by contracting slices in parallel.
We used it with KaHyPar as driver and optuna~\cite{Akiba2019} as optimization library.
As a cost function, we used ``flops'' to meet the optimization goal of our methods.
For the basic slicing method (\texttt{slicing}), we chose the target number of slices a priori by a sweep over the values $\{4, 8, 16, 32, 64, 128, 256\}$, similar to the number of partitions examined with our methods.
For the advanced dynamic slicing method (\texttt{slicing\_reconf}), which iteratively refines the contraction tree after each slicing step, the target number of slices could not be set.
Instead, we conducted a sweep over possible target sizes for intermediate tensors of $\{2^{10}, 2^{18}, 2^{26}, 2^{34}, 2^{42}, 2^{50}, 2^{58}\}$.
We note that such a wide range was necessary because, for large circuits, smaller target sizes frequently prevented the method from producing a result within the allotted time.
The Cotengra methods were given the same $10$‑minute time limit, and we report the best result obtained within that budget.
The cost was calculated using \cref{eq:con_serial} on a sliced tensor network to get the cost of a single slice, and then multiplying it by the number of sequential rounds required when the slices are distributed evenly across $256$ ranks.

Run time experiments were conducted on the SuperMUC-NG cluster of the \gls{LRZ}.
The cluster has 48-core Intel Xeon Platinum 8174 nodes with the Skylake microarchitecture.
Each node has $\SI{768}{\giga\byte}$ of RAM\@.
The interconnect is an Intel OmniPath network with $\SI{100}{\giga\bit\per\second}$.

\subsection{Experiments}
\subsubsection{Investigating Correlation with Time-to-Solution}
As an initial experiment, we investigated how well the chosen metric correlates with the actual time-to-solution.
As the metric is used by the simulated annealing approaches for optimization, a high correlation with the actual time-to-solution is crucial for good optimization results and ensuring that it is a good metric for comparison between methods.
Here, we ran the three partitioning-based methods for some of the circuits and then contracted the circuits with the resulting contraction paths on actual hardware.

\begin{figure}[htbp]
    \centering
    \subfloat[Theoretical computational cost compared to actual time-to-solution.]{\label{subfig:scatter}\includegraphics{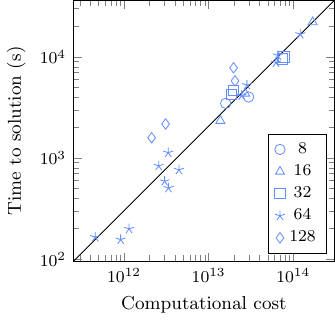}}
    \hfill
    \subfloat[Percentage of total run time spent in MPI communication versus total run time.]{\label{subfig:scatter_comm}\includegraphics{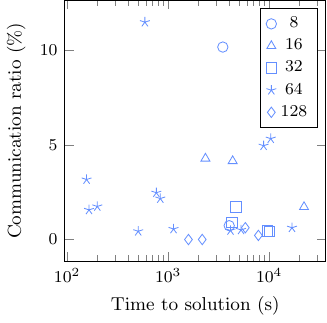}}
    \caption{Run time results for different circuit contractions with varying numbers of partitions.}
\end{figure}

\Cref{subfig:scatter} displays the theoretical cost metric or computational cost on the x-axis compared against the actual time-to-solution on the y-axis.
There is a clear dependence between the two variables, indicating that our chosen metric is a reasonable candidate for predicting time-to-solution, at least on the used hardware.

We additionally investigated the communication overhead, since we previously argued that it is often negligible in practice.
Specifically, we measured the time spent on raw data transmission, excluding idle time caused by load imbalance.
Such imbalance is inherent to tensor network contraction: the network structure can induce naturally uneven clustering, leading to tasks of differing computational weight.
While our method mitigates this imbalance where possible, it does so only when the total contraction cost is not increased.
\Cref{subfig:scatter_comm} shows the fraction of the run time spent on communication, which is calculated by determining the shorter of the time blocked for either send or receive MPI calls as this reflects the time spent in communication.
The communication time accounts for less than \SI{5}{\percent} in most cases.
The overhead is problem‑dependent, and we do not observe a consistent scaling trend with increasing problem size.

\subsubsection{Strong Scaling of Directed Simulated Annealing}

\begin{figure}[htbp]
    \centering
    \subfloat[qnn, 30 qubits.]{\includegraphics{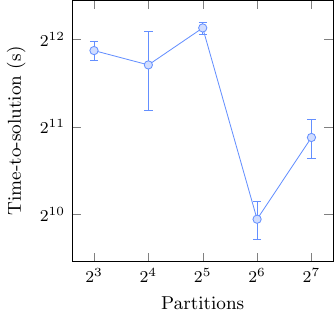}}
    \hfill
    \subfloat[twolocalrandom, 30 qubits.]{\includegraphics{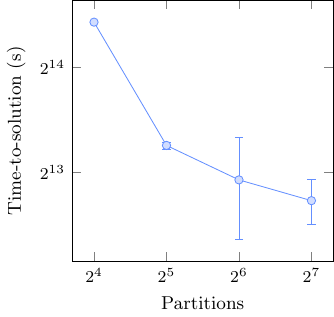}}
    \caption{Time-to-solution for increasing numbers of partitions after running directed simulated annealing with two different seeds.}
    \label{fig:strong_scaling}
\end{figure}

To investigate how our main method, directed simulated annealing, scales with increasing hardware resources, we evaluated the same problem when varying the number of partitions and the corresponding number of compute nodes.
We compare the results for two specific problems in MQT~Bench: \textit{qnn} and \textit{twolocalrandom}, as the former requires multiple all-to-all interactions while the latter uses only nearest-neighbor interactions, resulting in a sparser circuit structure.
The results are shown in \cref{fig:strong_scaling}.
While some problems can demonstrate a general improvement in time-to-solution with increased partitions, the overall observation is that time-to-solution does not decrease monotonically with increased partitions.
Beyond the inherent non-determinism of our method, arising from the random choices made during simulated annealing, this behavior indicates that the quality of a partitioning can depend strongly on the requested number of partitions.
Consequently, these results motivate performing an a priori sweep over candidate partition counts, rather than defaulting to the largest feasible number.

\subsubsection{Comparison of Methods}
\begin{figure}[htbp]
    \centering
    \includegraphics{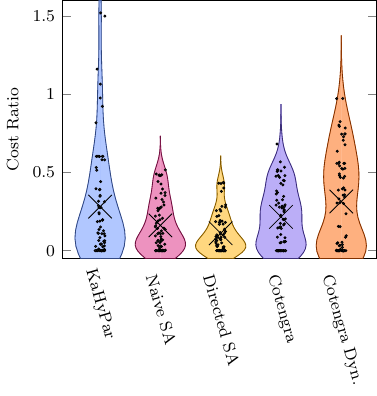}
    \hfill
    \includegraphics{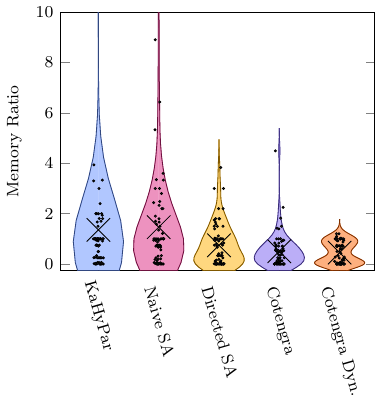}
    \caption{Theoretical computational and memory costs of all methods over all circuits, compared to a serial execution. The mean is indicated by an ``\texttimes'' marker. Lower values are better. KaHyPar exhibited one extreme outlier with a memory ratio of $31$, which lies outside the plotted axis range.}
    \label{fig:boxplot}
\end{figure}

We now performed a full sweep of the theoretical computational cost for each method on each circuit.
As a baseline, we used the cost of a serial contraction path found by the Greedy heuristic.
We divided the results from other methods by this baseline to obtain a problem-independent ratio, with lower values indicating greater improvement.
The results are aggregated in \cref{fig:boxplot}, which showcases the distribution of the computation ratios for all tested methods.
Although our primary objective is to reduce computational cost, we also report memory cost for practical relevance.
The results indicate that using simulated annealing to refine the generic partitioning found by KaHyPar reduced computational cost on average.
In particular, the outliers where the cost was worse than the serial contraction were mitigated.
It is also evident that guiding the simulated annealing algorithm outperformed the generic simulated annealing variant in several cases.
Additionally, although memory usage was not part of the optimization objective, the methods often reduced the required memory.
Interestingly, the basic Cotengra method reduced computational cost more than the dynamic Cotengra method, likely because the latter focused more on minimizing slicing overhead.

\begin{figure}[htbp]
    \centering
    \includegraphics{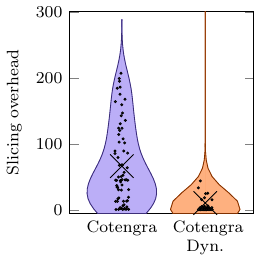}
    \caption{Slicing overhead of the Cotengra methods over all circuits. Cotengra Dynamic exhibited two extreme outliers at \num{463} and \num{62357}, which lie outside the plotted axis range.}
    \label{fig:slicing}
\end{figure}
\Cref{fig:slicing} shows the slicing overhead for the Cotengra methods, which we computed as the sum of operations of the sliced problem divided by the sum of operations of the serial baseline, similar to the definition in~\cite{Gray2021}.
Dynamic Cotengra achieved a lower slicing overhead compared to basic Cotengra, partly because it often sliced fewer legs.
However, in our setting, minimizing slicing overhead is not always helpful: since we assume a parallel contraction on $256$ compute nodes, any slicing overhead of $256$ or less does not increase the total time-to-solution.
This explains why basic Cotengra performed better on average, since its larger slicing overhead is better aligned with our parallel execution setting.

\begin{figure}[htbp]
    \centering
    \includegraphics{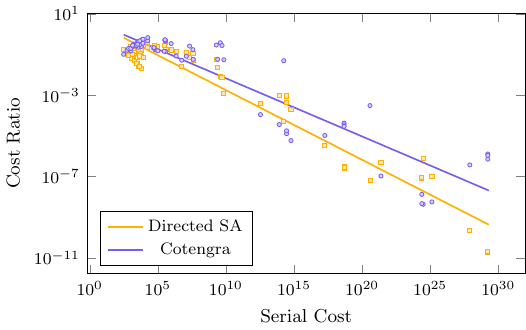}
    \caption{Comparison of the cost ratio of directed simulated annealing and basic Cotengra over the serial contraction cost.}
    \label{fig:problem_size}
\end{figure}
\Cref{fig:problem_size} shows the computational cost ratio of the directed simulated annealing and the basic Cotengra method over the serial contraction cost, which serves as a proxy for problem hardness.
In general, larger problems demonstrated greater potential for flop optimization by improving partitioning.
One possible explanation is that searching for higher‑quality contraction paths locally within partitions and then combining these localized paths is easier than directly searching for a single contraction path in the full, increasingly large tensor network with greedy heuristics.
Notably, Cotengra also utilizes partitioning for path finding, but not as a means of parallelizing the contraction task.

\begin{figure}[htbp]
    \centering
    \includegraphics{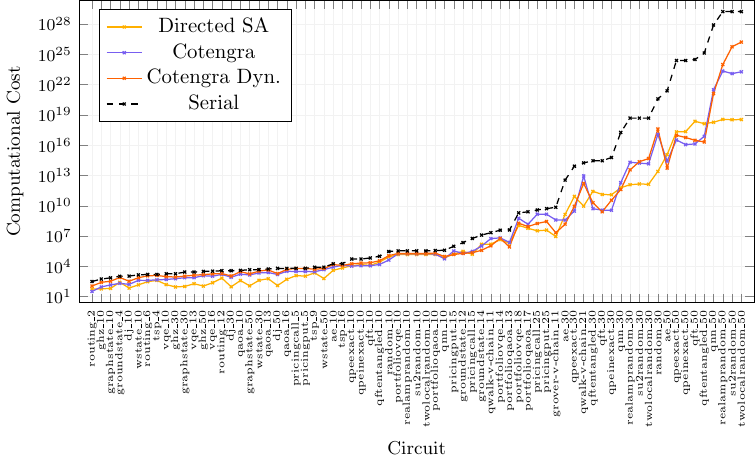}
    \includegraphics{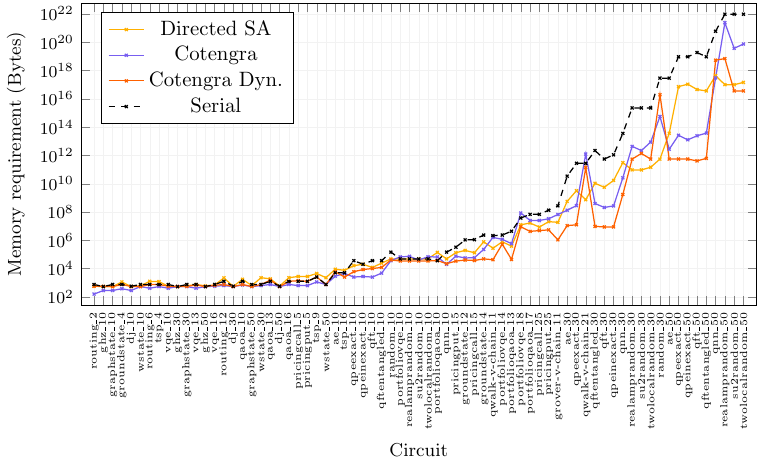}
    \caption{Comparison of the computational cost and memory requirements of the serial baseline, the directed simulated annealing method, and the Cotengra methods on all circuits, ordered by serial cost.}
    \label{fig:linegraph}
\end{figure}
Finally, we compared the absolute cost of the directed simulated annealing, basic Cotengra, dynamic Cotengra, and the serial baseline in \cref{fig:linegraph}.
All methods outperformed the serial baseline by several orders of magnitude.
Directed simulated annealing consistently achieved the lowest cost on small and medium‑sized circuits and remained competitive with both variants of Cotengra on larger instances.
There appears to be no clear classification of problem instances in which one method consistently outperforms the other; this suggests that multiple approaches should be considered when exploring contraction strategies for new tensor networks.
Overall, the directed simulated annealing method found the contraction path with the lowest computational cost for $47$ of the $71$ investigated circuits and was within \SI{10}{\percent} of the best result for $2$ additional circuits.
It reduced cost by up to $11$ orders of magnitude compared to the serial baseline.
For completeness, we also provide the actual computational cost and memory cost for all methods in \cref{appendix:results}.

\section{Conclusion}

In this work, we investigated the use of partitioning for distributed tensor network partitioning, with a focus on optimizing an initial partitioning using two simulated annealing methods for lower time-to-solution.
The methods were compared with two state-of-the-art HyperOptimizer methods provided by the Cotengra library by running them for the same amount of time and comparing the results.
We utilized operation count along the critical path as a metric and validated its accuracy by comparing it to the actual time-to-solution for a subset of problems, constrained by memory limitations.
While no single approach dominated for all circuits, our directed simulated annealing method achieved the lowest computational cost on the majority of instances and provided substantial gains on the hardest problems.
Future work could include finding a heuristic for a priori algorithm selection given a specific tensor network.

We note that in this work, we did not use slicing with our partitioning methods.
However, slicing could be applied on top of the partitioning to alleviate the memory restrictions seen in the found contraction trees.
Such an implementation would allow running larger problem sizes and open doors to comparison with additional state-of-the-art methods, such as those introduced by \cite{Pan2022b} and \cite{Gray2024}.
At the same time, the number of partitions should be decoupled from the available hardware resources to allow for greater flexibility.
In addition, our methods could be enhanced to also optimize for the ideal number of partitions to use, which is currently determined a priori by sweeping over plausible values.
Finally, the introduced methods could be integrated into other approaches that utilize tensor network partitioning, such as contraction path finding.

\section*{Acknowledgments}
We thank Marco De Pascale and Luigi Iapichino for their input about performance on the HPC system. We furthermore thank the Leibniz Rechenzentrum (LRZ) for the provided computation time on SuperMUC-NG.

\bibliographystyle{siamplain}
\bibliography{references}

\appendix
\newpage
\section{Full experimental results}
\label{appendix:results}
\mbox{}\vspace{\baselineskip}

\begin{footnotesize}
\setlength{\abovecaptionskip}{0pt}
\setlength{\belowcaptionskip}{0pt}
\setlength{\textfloatsep}{0pt}
\setlength{\intextsep}{0pt}
\begin{table}[H]
    \centering
    \caption{Comparison of theoretical computation cost of all methods.}\label{tab:flops}
\end{table}
\begin{longtable}{
        p{1.18cm}
        *{5}{S[round-mode=places,round-precision=2,scientific-notation=true,output-exponent-marker=\mathrm{e},table-format=1.2e2,detect-weight=true,]}
        *{2}{S[round-mode=places,round-precision=1,output-exponent-marker=\mathrm{e},table-format=5.1]}
}
        \toprule
        \textbf{File} & \multicolumn{5}{c}{\textbf{Computational Cost}} & \multicolumn{2}{c}{\textbf{Slicing Overhead}} \\
    \cmidrule(lr){2-6}\cmidrule(lr){7-8}
                      & {KaHyPar} & {Naive SA} & {Dir. SA} & {Cotengra} & {Cot. Dyn} & {Cotengra} & {Cot. Dyn}\\\midrule
ae10 & 7872.0 & 5544.0 & \bfseries 3980.0 & 8479.0 & 14692.5 & 46.00267076043069 & 1.2704810091488772\\
ae30 & 17793633248.0 & 8231472656.0 & 1369808256.0 & 379844876.0 & \bfseries 148241146.5 & 6.74428886146675 & 1.4457933351910448\\
ae50 & 6308281610509984.0 & 3748911077429136.0 & 1149666651179456.0 & 258947258961936.0 & \bfseries 54629277933924.5 & 1.870848976387508 & 1.4164329617870917\\
dj10 & \bfseries 70.0 & \bfseries 70.0 & \bfseries 70.0 & 162.0 & 330.0 & 114.67919075144508 & 1.0\\
dj30 & 150.0 & 150.0 & \bfseries 94.0 & 826.0 & 1098.0 & 184.8880070546737 & 1.0\\
dj50 & 166.0 & 166.0 & \bfseries 123.0 & 1546.0 & 1886.0 & 207.57797681770285 & 1.0\\
ghz10 & 62.0 & 62.0 & \bfseries 58.0 & 90.0 & 256.0 & 79.87593984962406 & 1.0\\
ghz30 & 116.0 & 116.0 & \bfseries 90.0 & 594.0 & 866.0 & 168.0883668903803 & 1.0\\
ghz50 & 188.0 & 188.0 & \bfseries 110.0 & 1154.0 & 1494.0 & 195.49334221038612 & 1.0\\
grphst10 & 91.0 & 91.0 & \bfseries 65.0 & 143.0 & 333.0 & 103.30860534124628 & 1.0\\
grphst30 & 135.0 & 135.0 & \bfseries 101.0 & 735.0 & 1055.5 & 175.8179245283019 & 1.0\\
grphst50 & 139.0 & 139.0 & \bfseries 116.0 & 1395.0 & 1777.0 & 199.6219101123596 & 1.0\\
grndst14 & 13234432.0 & 4559232.0 & 1523920.0 & 1027589.0 & \bfseries 386996.0 & 19.591511218677205 & 1.5965465145333284\\
grndst12 & 1141056.0 & 389272.0 & 304056.0 & \bfseries 190863.0 & 190964.0 & 13.554854531686912 & 1.26868129585662\\
grndst4 & 404.0 & 288.0 & 260.0 & \bfseries 205.0 & 767.0 & 43.02801358234296 & 1.0\\
grover11 & 300485312.0 & 224340264.0 & \bfseries 8965472.0 & 389918408.0 & 21057080.0 & 64.3507408808596 & 3.565324810760681\\
pfqaoa10 & 336160.0 & 174840.0 & \bfseries 146060.0 & 162806.0 & 205875.0 & 52.90291092716905 & 1.3894760380471731\\
pfqaoa13 & 9313344.0 & 3540272.0 & 1974028.0 & 2307144.0 & \bfseries 832549.0 & 35.1942813518062 & 44.39355027517932\\
pfqaoa17 & 493952928.0 & 183405816.0 & \bfseries 57844096.0 & 144900292.0 & 85791327.0 & 31.188977006517675 & 1.4727285369147998\\
pfvqe10 & 203936.0 & 165904.0 & \bfseries 145136.0 & 172581.0 & 188897.0 & 45.281373760074544 & 1.3041735708367854\\
pfvqe14 & 13075200.0 & 7158160.0 & \bfseries 4363904.0 & 6563784.0 & 5833317.0 & 37.3338612839495 & 1.4027258831323013\\
pfvqe18 & 1136822144.0 & 657903880.0 & \bfseries 113264912.0 & 574331099.0 & 186814217.0 & 14.412457775375236 & 1.510889597967155\\
prccll5 & 1744.0 & 1398.0 & \bfseries 1196.0 & 3042.0 & 6160.0 & 124.35323303457106 & 1.0\\
prccll15 & 1789696.0 & 293768.0 & \bfseries 156010.0 & 300304.0 & 220829.0 & 38.14326267714017 & 2.0683287267921564\\
prccll25 & 3747968032.0 & 1981833744.0 & \bfseries 32692732.0 & 1465019522.0 & 174627866.0 & 85.7781300707968 & 6.598415650175845\\
prcput5 & 1744.0 & 1314.0 & \bfseries 1064.0 & 3042.0 & 6160.0 & 124.35323303457106 & 1.0\\
prcput15 & 1433952.0 & 257556.0 & 162594.0 & 331547.0 & \bfseries 146799.0 & 46.4529477694942 & 2.5726421855979327\\
prcput25 & 426478176.0 & 353614432.0 & \bfseries 38327280.0 & 1428661194.0 & 273635932.0 & 102.0060878005015 & 3.784169366359131\\
qaoa10 & 425.0 & 425.0 & \bfseries 388.0 & 1650.5 & 2932.0 & 136.08215206185568 & 1.0\\
qaoa13 & 1048.0 & 856.0 & \bfseries 592.0 & 2424.5 & 4336.0 & 120.67692006269591 & 1.0\\
qaoa16 & 1152.0 & 880.5 & \bfseries 520.0 & 3211.0 & 4873.0 & 164.5382146439318 & 1.0\\
qft10 & 19232.0 & 13314.0 & 12176.0 & \bfseries 11524.0 & 22199.0 & 67.92809418486776 & 1.4434468017976425\\
qft30 & 2142507261312.0 & 613252685608.0 & 130841641008.0 & 3650030157.0 & \bfseries 2513938191.0 & 3.0214863342975056 & 4.278292017635973\\
qft50 & 1.9147797511792763e+19 & 6.400107702397722e+18 & 2.2992675294886492e+18 & \bfseries 1.3271407928526334e+16 & 3.0219185145608064e+16 & 1.1890604579989423 & 5.365522457917603\\
qftent10 & 81712.0 & 29120.0 & 26300.0 & \bfseries 15200.0 & 35665.0 & 45.91186286407767 & 1.3037166200018675\\
qftent30 & 3015170874016.0 & 1435829383496.0 & 264088592316.0 & \bfseries 5063329754.0 & 19641433727.0 & 2.673518213484394 & 15.80018822176798\\
qftent50 & 2.9871850037458e+19 & 6.618966309616047e+18 & 1.3499163063504082e+18 & 7.609464358074747e+16 & \bfseries 2.05765797712376e+16 & 1.0303527993111352 & 1.6331355815002435\\
qnn10 & 151360.0 & 102920.0 & 70200.0 & \bfseries 54930.0 & 91389.0 & 34.48440495553983 & 1.2597845665840843\\
qnn30 & 109394707959584.0 & 12497323736540.0 & 599803975328.0 & 1931944701656.0 & \bfseries 397431861097.5 & 2.441594697426287 & 18.71663569372216\\
qnn50 & 1.0109046350795227e+23 & 5.23981944444861e+21 & \bfseries 1.8272652117371735e+18 & 3.038207415287373e+21 & 1.265505401522656e+21 & 1.3023828623065894 & 1.967190657976836\\
qpe10 & 18320.0 & 13208.0 & 11350.0 & \bfseries 10520.0 & 17759.0 & 68.55911393713814 & 1.3818767812207398\\
qpe30 & 3350512850976.0 & 249226180568.0 & 84118916728.0 & \bfseries 2986166812.0 & 8804591074.0 & 3.707106151218314 & 25.020419906402783\\
qpe50 & 4.256554381502293e+19 & 6.990956837943908e+18 & 2.065050835140079e+17 & \bfseries 3.244842062242971e+16 & 9.50007189739008e+16 & 1.275262708450042 & 4.12506995456723\\
qpeix10 & 27776.0 & 18416.0 & 14784.0 & \bfseries 11508.0 & 20975.0 & 68.51839816719657 & 1.3865600017585418\\
qpeix30 & 1997158648640.0 & 396940623000.0 & 121292012336.0 & \bfseries 3541105493.0 & 35536655599.0 & 2.043645641354583 & 33.896619745588424\\
qpeix50 & 4.256554381502293e+19 & 4.872452309932536e+18 & 2.209354590930338e+17 & \bfseries 1.1346354436532824e+16 & 5.708278337079808e+16 & 1.182091937361689 & 3.878705990121407\\
qwalk11 & 32456544.0 & 9437442.0 & 1575472.0 & 5557103.5 & \bfseries 1078009.0 & 45.44894731011844 & 15.930058326877989\\
qwalk21 & 4447599141632.0 & 2968395504688.0 & \bfseries 9193088272.0 & 8789913662099.0 & 1615570680081.5 & 13.391387496649475 & 5.06833812149412\\
rnd10 & 338112.0 & 107982.0 & 85188.0 & \bfseries 42148.0 & 112983.0 & 30.4844700544671 & 1.3167356057835162\\
rnd30 & 1.2638226696124642e+17 & 2568194485399128.0 & \bfseries 25230801761564.0 & 1.1814169864145416e+17 & 3.942943413657004e+17 & 18.659007233661903 & 25.622894224193484\\
rlrnd10 & 203936.0 & 162912.0 & \bfseries 147768.0 & 174928.0 & 188897.0 & 45.04971287957441 & 1.3041735708367854\\
rlrnd30 & 121152994097412.0 & 56828602692880.0 & \bfseries 1207154271672.0 & 203719047977102.0 & 35513988636563.5 & 13.623888163274565 & 5.954356581652048\\
rlrnd50 & 6.956636233174756e+24 & 7.62390909861705e+22 & \bfseries 3.5674138526834545e+18 & 2.186549509574071e+23 & 9.60495145391602e+23 & 5.089015866261647 & 2.6749662774991068\\
rout2 & 98.0 & 84.0 & 56.0 & \bfseries 32.0 & 110.0 & 31.063106796116504 & 1.0\\
rout6 & 376.0 & 352.0 & \bfseries 300.0 & 402.0 & 1128.0 & 89.64450354609929 & 1.0\\
rout12 & 1000.0 & 1000.0 & \bfseries 674.0 & 1564.0 & 1913.0 & 86.49978298611111 & 1.7661997340332638\\
su2rnd10 & 203936.0 & 163872.0 & \bfseries 146628.0 & 171256.0 & 188897.0 & 51.0514791507518 & 1.3041735708367854\\
su2rnd30 & 121152994097412.0 & 49915273583868.0 & \bfseries 1428489586640.0 & 165794675183984.0 & 229033721159394.0 & 13.3703623566641 & 1.7637834797688998\\
su2rnd50 & 6.956636233174756e+24 & 1.9163822085514035e+23 & \bfseries 3.2988939340040223e+18 & 1.2238368130066838e+23 & 5.6090748718309295e+25 & 17.0762569795067 & 463.5860039644932\\
tsp4 & 620.0 & 512.0 & \bfseries 400.0 & 456.0 & 1244.0 & 89.79921875 & 1.0\\
tsp9 & 4840.0 & 3162.0 & \bfseries 2228.0 & 2902.0 & 5103.0 & 50.36144332699103 & 1.5734271369608588\\
tsp16 & 10592.0 & 7768.0 & \bfseries 6908.0 & 12427.0 & 12329.0 & 57.4298150556278 & 1.7276509372149769\\
twolr10 & 203936.0 & 164152.0 & \bfseries 145964.0 & 180490.0 & 188897.0 & 45.23516217777386 & 1.353782550648705\\
twolr30 & 121152994097412.0 & 51788725013712.0 & \bfseries 1340432455096.0 & 141094474941080.0 & 467693375418583.0 & 8.264977994277245 & 3.975974118187475\\
twolr50 & 6.956636233174756e+24 & 2.1168934324175874e+22 & \bfseries 3.448068464729884e+18 & 1.8562828448089197e+23 & 1.6277365807848566e+26 & 7.562270722972967 & 62357.86593409591\\
vqe10 & 340.0 & 260.0 & \bfseries 154.0 & 492.0 & 948.0 & 130.9694092827004 & 1.0\\
vqe13 & 406.0 & 406.0 & \bfseries 188.0 & 756.0 & 1314.0 & 143.09179104477613 & 1.0\\
vqe16 & 308.0 & 308.0 & \bfseries 234.0 & 1050.0 & 1768.0 & 147.35485651214128 & 1.0\\
wstate10 & 192.0 & 192.0 & \bfseries 144.0 & 404.0 & 740.0 & 108.11808510638298 & 1.0\\
wstate30 & 520.0 & 520.0 & \bfseries 406.0 & 2206.0 & 3324.0 & 159.74545970488083 & 1.0\\
wstate50 & 880.0 & 880.0 & \bfseries 612.0 & 4624.0 & 6076.0 & 186.4276025236593 & 1.0\\
\end{longtable}
\end{footnotesize}

\begin{footnotesize}
\setlength{\abovecaptionskip}{0pt}
\setlength{\belowcaptionskip}{0pt}
\setlength{\textfloatsep}{0pt}
\setlength{\intextsep}{0pt}
\addtocounter{table}{-1}
\begin{table}[H]
    \centering
    \caption{Comparison of theoretical memory cost of all methods.}\label{tab:memory}
\end{table}
\begin{longtable}{
        p{1.18cm}
        *{5}{S[round-mode=places,round-precision=2,scientific-notation=true,output-exponent-marker=\mathrm{e},table-format=1.2e2,detect-weight=true,]}
}
        \toprule
        \textbf{File} & \multicolumn{5}{c}{\textbf{Memory Requirement (Bytes)}} \\
    \cmidrule(lr){2-6}
                      & {KaHyPar} & {Naive SA} & {Dir. SA} & {Cotengra} & {Cot. Dyn}\\\midrule
ae10 & 10240.0 & 11264.0 & 9216.0 & \bfseries 2944.0 & 6144.0\\
ae30 & 603979776.0 & 537133056.0 & 570490912.0 & 134742016.0 & \bfseries 11010048.0\\
ae50 & 70368760954880.0 & 70368744833024.0 & 36833639538688.0 & 2753107591168.0 & \bfseries 554050789376.0\\
dj10 & 544.0 & 544.0 & 544.0 & \bfseries 288.0 & 544.0\\
dj30 & 576.0 & 576.0 & 560.0 & \bfseries 544.0 & \bfseries 544.0\\
dj50 & \bfseries 544.0 & \bfseries 544.0 & 592.0 & \bfseries 544.0 & \bfseries 544.0\\
ghz10 & 544.0 & 544.0 & 544.0 & \bfseries 288.0 & 560.0\\
ghz30 & \bfseries 544.0 & \bfseries 544.0 & \bfseries 544.0 & \bfseries 544.0 & \bfseries 544.0\\
ghz50 & \bfseries 544.0 & \bfseries 544.0 & 608.0 & \bfseries 544.0 & \bfseries 544.0\\
grphst10 & 576.0 & 576.0 & 544.0 & \bfseries 288.0 & 544.0\\
grphst30 & 576.0 & 576.0 & 608.0 & \bfseries 544.0 & 560.0\\
grphst50 & 576.0 & 576.0 & 640.0 & \bfseries 544.0 & \bfseries 544.0\\
grndst14 & 2359296.0 & 3178512.0 & 786432.0 & 229376.0 & \bfseries 49152.0\\
grndst12 & 270336.0 & 196608.0 & 196608.0 & 57344.0 & \bfseries 40960.0\\
grndst4 & 768.0 & 1152.0 & 1152.0 & \bfseries 384.0 & 768.0\\
grover11 & 536903680.0 & 203456512.0 & 18876544.0 & 67141632.0 & \bfseries 1089536.0\\
pfqaoa10 & 66560.0 & 328320.0 & 141312.0 & 67072.0 & \bfseries 37888.0\\
pfqaoa13 & 1179648.0 & 4196480.0 & 393216.0 & 589824.0 & \bfseries 45056.0\\
pfqaoa17 & 17039360.0 & 83902464.0 & 16810496.0 & 25690112.0 & \bfseries 4325376.0\\
pfvqe10 & 49152.0 & 49152.0 & 49152.0 & 68608.0 & \bfseries 34816.0\\
pfvqe14 & 2359296.0 & 786432.0 & 786432.0 & 1179648.0 & \bfseries 540672.0\\
pfvqe18 & 34603008.0 & 201345024.0 & 12582912.0 & 85065728.0 & \bfseries 9175040.0\\
prccll5 & 2560.0 & 2816.0 & 2816.0 & \bfseries 640.0 & 1408.0\\
prccll15 & 557056.0 & 229376.0 & 133120.0 & 59392.0 & \bfseries 37888.0\\
prccll25 & 2147614720.0 & 1142947968.0 & 8683520.0 & 25300992.0 & \bfseries 4915200.0\\
prcput5 & 2560.0 & 3584.0 & 2816.0 & \bfseries 640.0 & 1280.0\\
prcput15 & 327680.0 & 286720.0 & 133120.0 & 75776.0 & \bfseries 34304.0\\
prcput25 & 33816576.0 & 335609856.0 & 21012480.0 & 33603584.0 & \bfseries 5505024.0\\
qaoa10 & 2176.0 & 2176.0 & 1792.0 & \bfseries 704.0 & 768.0\\
qaoa13 & 2560.0 & 2432.0 & 1920.0 & \bfseries 768.0 & 1536.0\\
qaoa16 & 4224.0 & 4288.0 & 2304.0 & \bfseries 768.0 & 1280.0\\
qft10 & 12288.0 & 25728.0 & 12288.0 & \bfseries 2560.0 & 10240.0\\
qft30 & 274878169088.0 & 137573179392.0 & 5637144576.0 & 210763776.0 & \bfseries 8913152.0\\
qft50 & 5.76460752304472e+17 & 5.76460752304472e+17 & 4.503599628419072e+16 & 24464133718016.0 & \bfseries 412384002048.0\\
qftent10 & 36864.0 & 37376.0 & 24576.0 & \bfseries 4864.0 & 12288.0\\
qftent30 & 17196646400.0 & 103079256064.0 & 10747904000.0 & 402784256.0 & \bfseries 9699328.0\\
qftent50 & 1.4411518814296474e+17 & 1.4411518810954138e+17 & 3.602879705264947e+16 & 37658281639936.0 & \bfseries 618483679232.0\\
qnn10 & 147456.0 & 98304.0 & 49152.0 & \bfseries 20480.0 & 22528.0\\
qnn30 & 2199024304128.0 & 343600005120.0 & 309237647872.0 & 26105348096.0 & \bfseries 1745354752.0\\
qnn50 & 5.775866522102661e+17 & 7.493989779944977e+19 & 4.323455642275717e+17 & \bfseries 2.975190503836877e+17 & 5.332261958823444e+18\\
qpe10 & 9216.0 & 25728.0 & 16384.0 & \bfseries 2560.0 & 6144.0\\
qpe30 & 68786585600.0 & 86037757952.0 & 3221266432.0 & 285736960.0 & \bfseries 12582912.0\\
qpe50 & 5.764607523369779e+17 & 5.76460752303448e+17 & 7.205759403792806e+16 & 26456999067648.0 & \bfseries 549789892608.0\\
qpeix10 & 20480.0 & 49920.0 & 20480.0 & \bfseries 2816.0 & 8704.0\\
qpeix30 & 274878431232.0 & 34360295424.0 & 17179873280.0 & 272629760.0 & \bfseries 8945664.0\\
qpeix50 & 5.764607523369779e+17 & 1.4411518807641293e+17 & 1.0808639105689203e+17 & 12644383719424.0 & \bfseries 550292685056.0\\
qwalk11 & 8519680.0 & 1212416.0 & 280576.0 & 1622016.0 & \bfseries 43008.0\\
qwalk21 & 69793218560.0 & 51573161984.0 & \bfseries 738230272.0 & 1237000912896.0 & 137472508160.0\\
rnd10 & 135168.0 & 98304.0 & 49152.0 & \bfseries 36864.0 & 40960.0\\
rnd30 & 141836999983104.0 & 35184406167552.0 & \bfseries 549756370944.0 & 563224831328256.0 & 1.9140590474100736e+16\\
rlrnd10 & 49152.0 & 49152.0 & 49152.0 & 73728.0 & \bfseries 34816.0\\
rlrnd30 & 154618822656.0 & 2473901690880.0 & \bfseries 94489288704.0 & 4398113619968.0 & 550024253440.0\\
rlrnd50 & 1.475740229584206e+20 & 3.804641020179854e+19 & \bfseries 9.90791918021591e+16 & 2.3611832414348223e+21 & 6.922032627276841e+18\\
rout2 & 576.0 & 768.0 & 576.0 & \bfseries 160.0 & 576.0\\
rout6 & 768.0 & 1792.0 & 1280.0 & \bfseries 416.0 & 768.0\\
rout12 & 2304.0 & 2304.0 & 2240.0 & \bfseries 640.0 & 768.0\\
su2rnd10 & 49152.0 & 49152.0 & 49152.0 & 40960.0 & \bfseries 34816.0\\
su2rnd30 & 154618822656.0 & 558346272768.0 & \bfseries 94489288704.0 & 2201170739200.0 & 1377073889280.0\\
su2rnd50 & 1.475740229584206e+20 & 6.918091977602892e+18 & 9.90791918021591e+16 & 3.689348815171407e+19 & \bfseries 3.602879701896422e+16\\
tsp4 & 768.0 & 2304.0 & 1216.0 & \bfseries 544.0 & 768.0\\
tsp9 & 5120.0 & 9216.0 & 4608.0 & \bfseries 1152.0 & 2560.0\\
tsp16 & 12288.0 & 32960.0 & 7680.0 & 4608.0 & \bfseries 2560.0\\
twolr10 & 49152.0 & 49152.0 & 49152.0 & 69632.0 & \bfseries 34816.0\\
twolr30 & 154618822656.0 & 1382979600384.0 & \bfseries 146028896256.0 & 8796227239936.0 & 549890035712.0\\
twolr50 & 1.475740229584206e+20 & 5.764695569863803e+17 & 1.441151880758661e+17 & 7.378697629510664e+19 & \bfseries 3.602907189687091e+16\\
vqe10 & 1152.0 & 1216.0 & 576.0 & \bfseries 416.0 & 576.0\\
vqe13 & 2304.0 & 2304.0 & 768.0 & \bfseries 416.0 & 704.0\\
vqe16 & \bfseries 576.0 & \bfseries 576.0 & 704.0 & \bfseries 576.0 & 672.0\\
wstate10 & \bfseries 544.0 & \bfseries 544.0 & 608.0 & \bfseries 544.0 & \bfseries 544.0\\
wstate30 & 1280.0 & 1280.0 & 2304.0 & \bfseries 640.0 & 768.0\\
wstate50 & 2560.0 & 2560.0 & 2304.0 & \bfseries 768.0 & \bfseries 768.0\\
\end{longtable}
\end{footnotesize}

\end{document}